\documentclass[prl,twocolumn]{revtex4}%
\usepackage{amsmath}
\usepackage{amssymb}
\usepackage{amsfonts}
\usepackage{graphicx}%
\providecommand{\U}[1]{\protect\rule{.1in}{.1in}}

\begin{document}
\title{Dynamics of photoinduced Charge Density Wave-metal phase transition in
K$_{0.3}$MoO$_{3}$}
\author{A. Tomeljak$^{1,2}$, H. Sch\"{a}fer$^{1}$, D. St\"{a}dter$^{1}$, M.
Beyer$^{1}$, K. Biljakovic$^{3}$,\ and J. Demsar$^{1,2}$}
\affiliation{$^{1}$Physics Department and Center for Applied Photonics, Universit\"{a}t
Konstanz, D-78457, Germany}
\affiliation{$^{2}$Complex Matter Department, Jozef Stefan Institute, SI-1000, Ljubljana, Slovenia}
\affiliation{$^{3}$Institute of Physics, Zagreb, Croatia}

\begin{abstract}
We present first systematic studies of the photoinduced phase transition from
the ground charge density wave (CDW) state to the normal metallic (M) state in
the prototype quasi-1D CDW system K$_{0.3}$MoO$_{3}$. Ultrafast non-thermal
CDW melting is achieved at the absorbed energy density that corresponds to the
electronic energy difference between the metallic and CDW states. The results
imply that on the sub-picosecond timescale when melting and subsequent initial
recovery of the electronic order takes place the lattice remains unperturbed.

\end{abstract}
\maketitle

In superconductors (SC), it has been known for decades that an intense laser
pulse can non-thermally destroy the superconducting ground state
\cite{Testardi}. The energy required to destroy superconductivity should, in
case all the absorbed optical energy is kept in the electronic subsystem
during the process of superconductivity suppression, be equal to the
condensation energy (energy difference between the free energy of the SC and
normal states at T = 0 K). In recent experiments on MgB$_{2}$ \cite{MgB2} and
the high temperature superconductor La$_{2-x}$Sr$_{x}$Cu$_{2}$O$_{4}$
\cite{LSCOKusar} it has been shown, however, that the absorbed optical energy
required to suppress superconductivity is substantially higher than the
thermodynamically measured condensation energy \cite{LSCOKusar}. This
discrepancy was accounted for by considering in detail all energy relaxation
pathways on the timescale when superconductivity suppression is achieved. From
this analysis it follows that on this timescale a quasi-equilibrium between
the density of quasiparticles and high frequency phonons is achieved with most
of the absorbed energy density being stored in the phonon subsystem
\cite{MgB2,LSCOKusar}.

Charge density wave systems present another broken symmetry ground state. Here
upon cooling through the CDW transition temperature the translational symmetry
is broken \cite{Gruner}. The appearance of the long-range charge density
modulation is accompanied by the appearance of the gap in the single particle
excitation spectrum at the Fermi level, while the collective excitations of
the CDW state are the so called amplitude and phase mode \cite{Gruner}. While
real-time studies of photoexcited quasiparticle and collective mode dynamics
in CDW compounds have been quite extensive in the weak and moderate
perturbation regime \cite{BBprl,TaS2,Dvorsek,Perfetti,NbSe3}, and a reasonable
understanding of the underlying relaxation processes has been obtained,
systematic studies in the high perturbation regime, where the energy of the
optical excitation pulse is enough to drive the phase transition from the CDW
ground state to the normal metallic state \cite{Sagar,Schmitt} are still lacking.

In this Letter we report on the first systematic study of carrier and
collective mode dynamics in a prototype quasi-1 dimensional CDW K$_{0.3}%
$MoO$_{3}$. Systematic temperature and excitation density dependent
measurements of the photoinduced reflectivity changes reveal that the phase
transition from the ground CDW state to the normal metallic state can be
achieved on the femtosecond timescale. From the energy conservation law it
follows that the phase transition is non-thermal in origin; i.e. the phase
transition is not a result of a simple heating of the sample to above the
critical temperature. The absorbed energy density required to optically induce
the phase transition is found to be comparable to the electronic energy
difference upon CDW condensation. These results give new insight in the
ultrafast processes governing the relaxation dynamics in low dimensional CDW
systems. In particular, the results suggest that on the timescale shorter than
the period of the characteristic lattice vibrations ($\approx0.6$ ps in
K$_{0.3}$MoO$_{3}$, which is the inverse frequency of the amplitude mode
\cite{BBprl,Travaglini}) the charge density modulation is suppressed while the
lattice remains unperturbed keeping the 2 k$_{F}$ modulation.

In the experiments described here, we utilized a degenerate optical pump-probe
technique to study the excitation intensity and temperature dependence of the
photoinduced reflectivity dynamics in single crystals of blue bronze K$_{0.3}%
$MoO$_{3}$. We used a commercial Ti:Sapphire amplifier producing 6 $\mu$J,\ 50
fs laser pulses at $\lambda$ = 800 nm (photon energy of 1.55 eV) at a variable
repetition rate between 9 and 250 kHz. The laser was used as a source of both
excitation and probe pulses. Samples were mounted in an optical helium flow
cryostat, with both excitation and probe beam entering the sample at near
normal incidence. Due to the strong anisotropy of the induced changes in
reflectivity with respect to light polarization \cite{BBprl}, the probe laser
beam was polarized along the chain [010] direction, while the excitation beam
was polarized along the perpendicular [102] direction \cite{Crystal}. To
ensure a homogeneous excitation profile, the diameter of the pump beam at the
sample position was twice the diameter of the probe beam. To determine the
photoexcitation density at the position of the sample with high precision we
used a beam profiler. We fitted the beam profile with a Gaussian, and the
excitation fluences, $F$, used throughout the paper correspond to the maximum
fluence at the center of the beam. Low thermal conductivity in K$_{0.3}%
$MoO$_{3}$ \cite{cpKwok} can lead to a pronounced increase of the equilibrium
temperature in the probed volume, proportional to the average laser power.
Therefore, the high excitation experiments were performed at a low repetition
rate (10-30 kHz), where at $F=1$ mJ/cm$^{2}$ the temperature increase is
$\lesssim10$ K.

\begin{figure}[h]
\begin{center}
\includegraphics[width=8.2263cm]{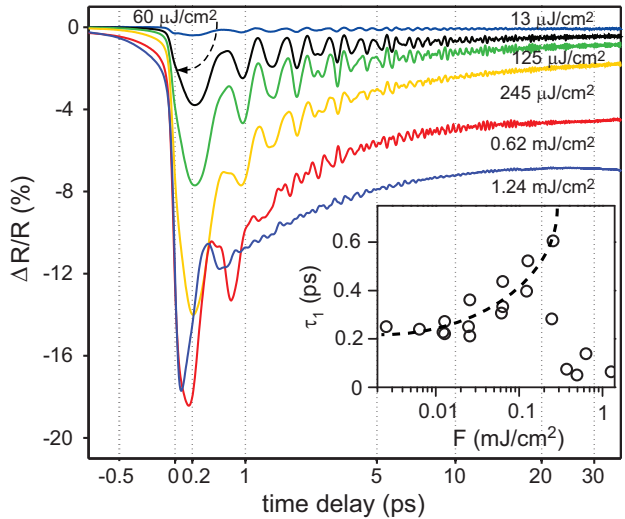}
\end{center}
\caption{Photoinduced reflectivity changes in K$_{0.3}$MoO$_{3}$ at 4 K
following photoexcitation with 50 fs optical pulses at different excitation
fluences \cite{shift}. Inset: the $F-$dependence of the initial decay time
$\tau_{1}$ determined by fitting the data as described in Ref.\cite{Tomeljak}.
$\tau_{1}(F)$ displays critical behavior (dashed line is a guide to the eye)
near the threshhold fluence of $\approx200$ $\mu$J/cm$^{2}$. }%
\end{figure}

Figure 1 presents the induced reflectivity transients taken at 4 K at several
excitation densities. At low excitation densities the data show the same
behavior as previously reported \cite{BBprl}. The decay dynamics of the
(incoherent) electronic response show a bi-exponential decay with timescales
$\tau_{1}\approx0.3$ ps and $\tau_{2}\approx7$ ps. The former one, showing
critical slowing down upon approaching T$_{c}^{3D}$, was attributed to the
recovery of the CDW gap, while the second one was initially tentatively
attributed to an overdamped phase mode \cite{BBprl}. Recent detailed studies
of the dynamics as a function of $F$ and applied external electric field
however suggest that this longer timescale more likely presents the second
stage of the CDW recovery \cite{Tomeljak}. On top of the incoherent transient
an oscillatory (coherent) signal is observed whose Fourier transform, obtained
by Fast Fourier Transform (FFT) analysis, shows several frequency components
which can be attributed to the coherently excited amplitude mode (the
strongest mode at 1.68 THz) and several other phonon modes
\cite{BBprl,Sagar,Tomeljak}.

While the photoinduced transient is linear in $F$ over several orders of
magnitude \cite{Tomeljak}, we observe pronounced changes upon increasing the
excitation intensity into the 100 $\mu$J/cm$^{2}$ range. The electronic
component shows clear saturation at $F\gtrsim200$ $\mu$J/cm$^{2}$ with a
maximum induced change in reflectivity approaching 18 \% \cite{comment}. One
is tempted to ascribe this saturation behavior in reflectivity change to the
photoinduced suppression of the CDW order, where this difference in
reflectivity corresponds to the change in reflectivity between the normal
metallic and the CDW state. Since no measurement of the temperature dependence
of reflectivity at optical frequencies is reported to date, we have performed
thermomodulation measurements to determine the magnitude and sign of change in
equilibrium reflectivity upon increasing the temperature to above T$_{c}^{3D}%
$. We measured the reflectivity difference between CDW and metallic state at
1.55 eV (800 nm) by heating the sample to above the phase transition using a
CW laser. The data show that the reflectivity at 800 nm indeed decreases upon
increasing the temperature. The change in reflectivity of $\sim10$ \% was
observed upon heating from 160 K to just above T$_{c}^{3D}$. At temperatures
above T$_{c}^{3D}$ the corresponding change in equilibrium reflectivity was
less than 1 \%. Therefore the observed reflectivity change and its saturation
behavior upon increasing the excitation fluence are consistent with the
photoinduced CDW-M phase transition.

\begin{figure}[h]
\begin{center}
\includegraphics[width=8.2cm]{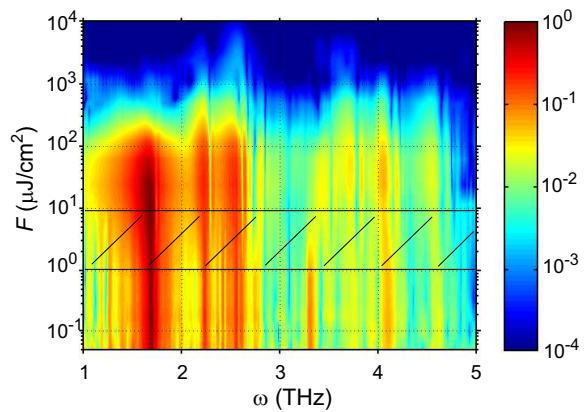}
\end{center}
\caption{$F$-dependence of the Fourier transform of the oscillatory signal at
4 K. The data were, prior to FFT analysis (time window of 50 ps), normalized
to $F$ for clarity. The area between 1.1 and 9 $\mu$J/cm$^{2}$ is obtained by
interpolation between the low $F$ data recorded with the Ti:sapphire
oscillator and the data recorded with the amplified system. Color coding
represents the FFT amplitude. Strong suppression of the AM ($\approx1.68$ THz)
is observed at $F\approx200$ $\mu$J/cm$^{2}$, while the other modes vanish at
fluences that are about one order of magnitude higher.}%
\end{figure}

From the rise-time of the reflectivity transient it also follows that this
transition happens on the 100 fs timescale after photoexcitation (rise-time is
becoming shorter upon increasing $F$). The initial decay time, $\tau_{1}$,
shows a pronounced increase near the threshold fluence, followed by a rapid
drop as shown in inset to Fig. 1 (the secondary decay time $\tau_{2}$ shows
only a slight decrease upon increasing $F$). This critical slowing down of
relaxation below the threshold for the CDW melting shows similar behavior as
in the weak perturbation limit upon increasing the temperature towards
T$_{c}^{3D}$ \cite{BBprl}.

Additional support to the assignment of the saturation behavior to the
photoinduced CDW-M transition comes from the study of the oscillatory
response, shown in Fig. 2. Here clear suppression of the amplitude mode, which
presents a fingerprint of the CDW state, is observed at comparable fluences.
Figure 2 shows the 2 dimensional surface plot of the FFT spectrum of the
oscillatory signal as a function of excitation intensity over several orders
of magnitude. Several sharp lines are observed, the strongest being that of
the amplitude mode (AM) at about 1.68 THz (56 cm$^{-1}$). The two second
strongest modes at 2.25 THz (74 cm$^{-1}$) and 2.55 THz (85 cm$^{-1}$)
correspond to zone-folding modes \cite{SagarRaman,Pouget}. In addition, some
weaker modes in the range between 3.5-4.5 THz are also observed as in Raman
\cite{SagarRaman}. We should also mention the weak side modes observed in the
vicinity of the AM as well as the two zone-folding modes. These are not
artefacts due to FFT. From the ratio of their amplitudes with respect to the
main peaks, and from the considerably weaker damping constants compared to the
main modes we ascribe these to their corresponding surface modes.%

\begin{figure}
[ptb]
\begin{center}
\includegraphics[
trim=2.229774in 4.186314in 2.350751in 4.231707in,
height=7.5125cm,
width=8.5229cm
]%
{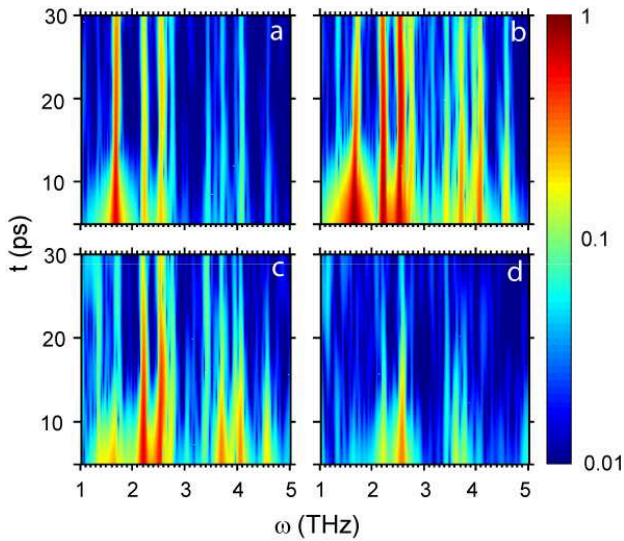}%
\caption{The time evolution of the FFT spectrum at 4K recorded at $F=12$ $\mu
$J/cm$^{2}$, 130 $\mu$J/cm$^{2}$, \textbf{ }0.38 mJ/cm$^{2}$, and \textbf{
}2.5 mJ/cm$^{2}$ (panels a-d). The plots are obtained by time-windowed FFT
analysis with a 5 ps time window. The AM is strongly suppressed at $F\geq200$
$\mu$J/cm$^{2}$, while the high frequency modes are present up to $F\approx3$
mJ/cm$^{2}$. }%
\end{center}
\end{figure}

While Fig. 2 clearly shows strong suppression and increased damping of the AM
above $F\approx200$ $\mu$J/cm$^{2}$, suppression of other phonons follow at
the excitation densities that are about one order of magnitude higher in $F$.
More information about the nature of the photoinduced CDW-M phase transition
can be gained by looking at the evolution of the FFT spectrum with time at
different $F$, which is presented in Fig. 3 for four different fluences. Here
one clearly sees that the AM vanishes at $F$ $\gtrsim300$ $\mu$J/cm$^{2}$
while the two zone-folding modes persist up to $F$ $\gtrsim2.5$ mJ/cm$^{2}$
finally disappearing above $F$ $\gtrsim3$ mJ/cm$^{2}$. The fact that high
frequency modes survive above the threshold\ fluence for the photoinduced
CDW-M phase transition suggests that on a short timescale time following the
photoinduced CDW-M transition and subsequent recovery on the sub-ps timescale
(inset to Fig. 1) the lattice remains largely unperturbed. In other words,
photoexcitation with a 50 fs optical pulse of $F>F_{sat}$ induces melting of
the electronic density modulation, which partially recovers on the sub-ps
timescale, while the lattice is - on this timescale - uncoupled from the
electron subsystem and retains its 2k$_{F}$ modulation. Only in the second
step of relaxation, which proceeds on the 10 ps timescale, the CDW can be
described with a single order parameter where electrons adiabatically follow
the lattice.

\begin{figure}[h]
\begin{center}
\includegraphics[width=8.0616cm]{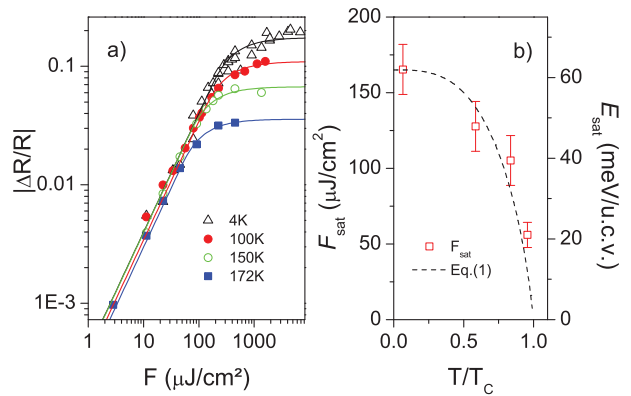}
\end{center}
\caption{a) Excitation fluence dependence of the induced reflectivity maximum
at several temperatures below T$_{c}^{3D}$. The solid lines are fits with the
simple saturation model \cite{LSCOKusar}. b) Saturation fluences extracted
from a) and the corresponding absorbed energy densities in meV per unit cell
volume. The dashed line represents the calculated T-dependence of $E_{sat}$
given by Eq.(1), where BCS T-dependence of $\Delta$ was used.}%
\end{figure}

We studied the $F$-dependence of the amplitude of the electronic signal at
different temperatures below T$_{c}^{3D}$ (Fig. 4). Upon increasing the
temperature saturation appears at decreasing values of $F$. In order to
determine the temperature dependence of the saturation fluence, $F_{sat}$, we
fit the data with the simple saturation model, which is described in detail in
Ref. \cite{LSCOKusar}. The temperature dependence of $F_{sat}$, together with
the corresponding absorbed energy density per unit cell volume \cite{Esat},
$E_{sat}$, is presented in Fig. 4b. The first question is whether melting of
the CDW state is a result of simple pulsed heating of the excited volume or is
it non-thermal in origin? We calculated the expected temperature rise $\Delta
T$ which corresponds to $E_{sat}$, using $E_{sat}=%
{\textstyle\int\nolimits_{T_{0}}^{T_{0}+\Delta T}}
c_{p}(T)dT$. Here $c_{p}$ is the total specific heat \cite{cpKwok}. At
$T_{0}=4$ K and $E_{sat}\approx60$ meV/u.c.v. we obtain $\Delta T$
$\lesssim40$ K, while at higher $T_{0}$ this value is considerably smaller
($\Delta T$ $\lesssim3$ K at 172 K and $E_{sat}\approx20$ meV/u.c.v.). In
fact, the absorbed energy density required to heat K$_{0.3}$MoO$_{3}$ from 4K
to its phase transition T$_{c}^{3D}$ = 183K is about $600$ meV/u.c.v., an
order of magnitude higher than $E_{sat}$. It follows that the photoinduced
CDW-M phase transition is non-thermally driven.

From the observation of coherently excited zone-folded modes at fluences up to
one order of magnitude higher than $F_{sat}$ it follows, that on the sub-ps
timescale after photoexcitation, the electrons are nearly uncoupled from the
lattice. In this case $E_{sat}$ should be compared to the energy gain of the
electronic subsystem upon CDW condensation, $E_{el}$. To estimate $E_{el}$ we
used the mean-field expression in the weak coupling limit, given by Eq. 3.40
in Ref. \cite{Gruner}:%
\begin{equation}
E_{el}=-n(\epsilon_{F})\Delta^{2}\left(  \frac{1}{2}+\log\frac{2\epsilon_{F}%
}{\Delta}\right)  .
\end{equation}
Here $n(\epsilon_{F})$ is the normal state density of states at the Fermi
energy, $\epsilon_{F}$, and $\Delta$ is the value of the CDW gap. Using
$\Delta$ = 60 meV, $\epsilon_{F}=0.24-0.39$ eV and $n(\epsilon_{F})$ $=4-6$
eV$^{-1}/$u.c.v. \cite{Gruner,PauliSus} we obtain $E_{el}(4$K$)=37-66$
meV/u.c.v.. This value is in excellent agreement with $E_{sat}(4$K$)\approx60$
meV/u.c.v., giving further support to the argument that during photoinduced
CDW melting the electronic order parameter is decoupled from the lattice on
the sub-ps timescale. At $F\gtrsim3$ mJ/cm$^{2}$, which corresponds to the
absorbed energy density of 1 eV/u.c.v., other modes are also completely
suppressed. This energy density is in good agreement with the calculated
energy density required to heat up the excited volume to above T$_{c}^{3D}$.
In this regime, recovery proceeds on a much longer timescale which is
determined by heat diffusion out of the excited volume. The temperature
dependence of $E_{sat}$, shown in Fig. 4b, shows good agreement with the
expected T-dependence of $E_{el}$. The dashed line in Fig. 4b shows
$E_{el}(T)$ calculated from Eq.(1), where the BCS T-dependence of $\Delta$ was
used \cite{Gruner}.

We have shown that in K$_{0.3}$MoO$_{3}$ the photoinduced CDW-M phase
transition is non-thermal and takes place on the 100 fs timescale. The good
agreement between measured $E_{sat}$ and calculated $E_{el}$, the observation
of the order parameter recovery on the sub-ps timescale, and the observation
of zone-folded phonons high above $E_{sat}$ suggest that during the process of
melting and sub-ps recovery of the electronic modulation the lattice remains
nearly frozen. This has an important implication for understanding the
ultrafast relaxation processes in systems with reduced dimensionality, in
particular for the systems with strong electron-phonon interactions that lead
to phenomena like charge density modulation. The initial reconstruction of the
CDW state is found in all systems studied thus far to proceed on the sub-ps
timescale \cite{BBprl,TaS2,Dvorsek,Perfetti,NbSe3,Sagar,Schmitt}. Importantly,
this timescale is one to two orders of magnitude faster than in the
high-T$_{c}$ superconductors \cite{LSCOKusar}, and is indeed close to the
typical timescale for electron-phonon thermalization. The formation of the CDW
requires freezing of a phonon and our results do imply that the lattice
remains frozen in its modulated state on the sub-ps timescale after
perturbation. Therefore, the extremely fast order parameter recovery in this
entire class of low-dimensional materials
\cite{BBprl,TaS2,Dvorsek,Perfetti,NbSe3,Sagar,Schmitt} could be a consequence
of the fact that on the short timescale after photoexcitation the lattice
remains in its unperturbed state. Thereby, the retaining 2$k_{F}$ modulation
presents a strong potential well driving ultrafast reformation of the charge
density modulation. Clearly, further theoretical studies as well as studies of
the ultrafast structural dynamics are required to shed additional light on
these fascinating phenomena.

\begin{acknowledgments}
We would like to acknowledge V.V. Kabanov, P. van Loosdrecht, T. Dekorsy, and
L. Degiorgi for valuable discussions. This work was supported by
Sofja-Kovalevskaja Grant from the Alexander von Humboldt Foundation, Center
for Applied Photonics and Zukunfts Kolleg at the University of Konstanz,
Croatian MSES project No. 035-0352827-2842 and AdFutura.
\end{acknowledgments}


\begin{thebibliography}{99}
\bibitem {shift}For presentation purposes x-axis is logarithmic with zero time
delay shifted to 1.5 ps prior to logarithmation.
\bibitem {comment}Similar saturation behavior was observed also in Ref.
\cite{Sagar}. There saturation is seen at higher fluences, which can be
ascribed to the different experimental configuration used.
\bibitem {Esat}Absorbed energy density is calculated using published values
\cite{Degiorgi} of the dielectric constant at 1.55 eV.
\end{thebibliography}
\end{document}